\documentclass[fleqn,usenatbib]{mnras}

\usepackage{newtxtext,newtxmath}

\usepackage[T1]{fontenc}

\DeclareRobustCommand{\VAN}[3]{#2}
\let\VANthebibliography\thebibliography
\def\thebibliography{\DeclareRobustCommand{\VAN}[3]{##3}\VANthebibliography}

\usepackage{graphicx}	
\usepackage{amsmath}	
\usepackage[dvipsnames]{xcolor}

\newcommand{\MSUN}{{\rm M}_{\sun}}

\newcommand{\RHOST}{r_{\rm 200c,\,host}}
\newcommand{\MSTAR}{M_{\rm *}}
\newcommand{\MGEN}[1]{M_{\rm #1}}
\newcommand{\RGEN}[1]{R_{\rm #1}}
\newcommand{\XXC}{Smallest Merger B}
\newcommand{\XXCV}{Smaller Merger B}
\newcommand{\XC}{Fiducial}
\newcommand{\XCX}{Larger Merger B}
\newcommand{\XCXX}{Largest Merger B}

\title[VINTERGATAN-GM: Impact of mergers on satellites]{VINTERGATAN-GM: How do mergers affect the satellite populations of MW-like galaxies?}

\author[G. D. Joshi et al.]{
Gandhali D. Joshi$^{1}$\thanks{E-mail: g.joshi@ucl.ac.uk},
Andrew Pontzen$^{1}$,
Oscar Agertz$^{2}$,
Martin P. Rey$^{3}$,
Justin Read$^{4}$
and Florent Renaud$^{2,5,6}$
\\
$^{1}$Department of Physics and Astronomy, University College London, Gower St., London WC1E 6BT, UK \\
$^{2}$Lund Observatory, Department of Astronomy and Theoretical Physics, Lund University, Box 43, SE-221 00 Lund, Sweden\\
$^{3}$Sub-department of Astrophysics, University of Oxford, DWB, Keble Road, Oxford OX1 3RH, UK\\
$^{4}$Department of Physics, University of Surrey, Guildford GU2 7XH, UK\\
$^{5}$Observatoire Astronomique de Strasbourg, Universit\'e de Strasbourg, CNRS UMR 7550, F-67000 Strasbourg, France\\
$^{6}$University of Strasbourg Institute for Advanced Study, 5 all\'ee du G\'en\'eral Rouvillois, F-67083 Strasbourg, France
}

\date{Accepted XXX. Received YYY; in original form ZZZ}

\pubyear{2024}

\begin{document}
\label{firstpage}
\pagerange{\pageref{firstpage}--\pageref{lastpage}}
\maketitle

\begin{abstract}
We investigate the impact of a galaxy's merger history on its system of satellites using the new \textsc{vintergatan-gm} suite of zoom-in hydrodynamical simulations of Milky Way-mass systems. The suite simulates five realizations of the same halo with targeted `genetic modifications' (GMs) of a $z \approx 2$ merger, but resulting in the same halo mass at $z=0$. We find that differences in the satellite stellar mass functions last for $2.25-4.25$ Gyr after the $z \approx 2$ merger; specifically, the haloes that have undergone smaller mergers host up to 60\% more satellites than those of the larger merger scenarios. However, by $z=0$ these differences in the satellite stellar mass functions have been erased. The differences in satellite numbers seen soon after the mergers are driven by several factors, including the timings of significant mergers (with $\MGEN{200c}$ mass ratios $>1:30$ and bringing in $\MSTAR \geq 10^{8}\MSUN$ at infall), the masses and satellite populations of the central and merging systems, and the subsequent extended history of smaller mergers. The results persist when measured at fixed central stellar mass rather than fixed time, implying that a host's recent merger history can be a significant source of scatter when reconstructing its dynamical properties from its satellite population.
\end{abstract}

\begin{keywords}
galaxies: dwarf -- galaxies: formation -- galaxies: evolution -- galaxies: interactions
\end{keywords}



\section{Introduction}
The $\Lambda$CDM cosmological model predicts that structure formation in the Universe is hierarchical -- smaller dark matter (DM) haloes are formed at earlier times and eventually coalesce to form larger structures, subsequently resulting in the merging of the galaxies formed within such haloes \citep{White1978,Peebles1980,Lacey1993}. The merger history of a galaxy plays a vital role in determining several of its properties such as its star-formation history, stellar composition and morphology. It follows that the large diversity of galaxy properties found in the Universe is at least partly driven by the vast range of merger histories galaxies undergo.

The majority of galaxies in the Universe are found in dense environments, either as part of groups and clusters of similar or more massive galaxies, or surrounded by their own system of lower mass (i.e. dwarf) galaxies \citep{Eke2004}. The number and properties of such satellite galaxies within a system are expected to be intrinsically tied to the precise assembly history of the overall dark matter (DM) halo it is embedded in \citep[e.g. see][and references therein]{Bose2020,Smercina2022}, as well as the interactions between the satellite galaxies and their environment. While environmental processes play a crucial role in determining satellite properties such as colour, star-forming status and morphology \citep[e.g.][]{Dressler1980,Balogh2004,Hogg2004,Kauffmann2004,Blanton2005,Wetzel2012,Alpaslan2015}, the overall census of the satellite population in a system and the kinematics of the satellite ensemble is primarily dependant on the mass of the host halo and its assembly and merger history \citep[e.g see][]{Giocoli2010}.

Dwarf galaxies in particular are an interesting mass regime in which to test models of galaxy formation. Their low masses and shallow potentials make them more responsive to both internal evolutionary processes as well as environmental factors, although their faintness presents significant challenges to their observations. Nonetheless, over the last few decades the list of observed dwarfs has grown from those around the Milky Way (MW) and Andromeda (M31) and more broadly within the Local Group (LG) \citep[e.g.][]{McConnachie2012,Martin2016,McConnachie2018} to other nearby galaxies \citep{Martin2013,Mueller2015,Carlin2016,Smercina2018,Crnojevic2019,Bennet2020}, as well as broader surveys such as SAGA \citep{Geha2017,Mao2021} and ELVES \citep{Carlsten2022}. These observations have shown that while MW-mass hosts exhibit a large diversity in satellite populations, the MW satellite populations are broadly consistent with those of other similar-mass hosts.

Satellite abundances around MW-mass hosts can provide an important benchmark for models of structure formation as well as galaxy formation. In the past decades, several observational studies have aimed at understanding the expected satellite abundances and their diversity in such hosts. Within the SAGA survey (Stage II), the 36 MW-mass hosts have a wide range in richness and luminosity functions (LFs) and the MW satellite LF is shown to be consistent within this range \citep{Mao2021}. Additionally, they find that the total number of satellites (with $M_{r,0}<-12.3$) appears to be positively correlated with the \textit{K}-band luminosity of the central galaxy, as well as the \textit{r}-band luminosity of the most massive satellite in the system. The former correlation likely reflects the increased number of satellites in more massive systems, while the latter may hint towards more satellites being found in hosts which have had more recent merger histories, under the assumption that due to dynamical friction, more massive satellites coalesce with the central galaxy faster than less massive ones, resulting in larger magnitude gaps at longer time intervals after a merger event. The \citet{Carlsten2021} study of 30 MW-like hosts in the ELVES survey finds a similar correlation between satellite abundance and the host's \textit{K}-band luminosity, and \citet{Danieli2022} show a wide diversity in satellite mass functions (MFs) with the same dataset.

The merger history of a central galaxy will inevitably influence its satellite accretion history, which in turn may affect the present day properties of the satellite populations around such systems. Precisely how the merger histories affect the satellites however remains to be understood. \citet{Smercina2022} investigated the impact of merger histories on the satellites of MW mass systems with a compilation of observations of satellites around the MW, M31 and six other MW-mass galaxies. They also find a surprisingly tight positive correlation between the number of satellites in a system (with $M_{V}<-9$) and the stellar mass of of the most dominant merger experienced by the galaxy, which they define as the larger of the total accreted stellar mass or the mass of the most massive satellite within the system. With extended data from the SAGA survey (xSAGA), \citet{Wu2022} show that the number of satellites in a system increases with decreasing \textit{r}-band magnitude gap between the central galaxy and its most luminous satellite, implying that hosts with earlier accretion histories have fewer satellites at present day, again due to similar arguments as above for the correlation with the luminosity of the brightest satellite.

One of the key challenges to obtaining such results from observations is the uncertainty in reconstructing the merger and accretion histories of the hosts. Simulations on the other hand allow us direct access to this information and can provide important insights into the process of satellite accretion. There have been several efforts to study dwarf satellites around MW-like systems in simulations that have broadly been able to reproduce the satellite MFs consistent with that of the MW, with the exception of the most massive MW satellites that are not always recovered. These include zoom-in hydrodynamical simulations of MW-, M31- and LG-like hosts e.g. using the FIRE galaxy model \citep{Hopkins2014,Hopkins2017,GarrisonKimmel2019}, the LATTE simulation \citep{Wetzel2016Latte}, the APOSTLE suite \citep{Sawala2016Apostle}, the NIHAO simulations \citep{Wang2015Nihao,Buck2019}, the ARTEMIS simulations \citep{Font2020Artemis,Font2021} and the DC Justice League simulations \citep{Applebaum2021JusticeLeague}, along with large-volume simulations e.g. IllustrisTNG50 \citep{Pillepich2019TNG,Nelson2019TNG,Engler2021} and ROMULUS25 \citep{Tremmel2017Romulus25,VanNest2023}, as well as using semi-analytical models \citep[e.g.][]{Macchio2010,Li2010,Font2011,Starkenburg2013,Jiang2021SatGen}. Furthermore, \citet{Starkenburg2013,GarrisonKimmel2019,Font2021,Engler2021,VanNest2023} all find positive correlations (to varying degrees) between the number of satellites and one or more host properties including halo mass, central stellar mass or central \textit{K}-band luminosity, that are at least qualitatively consistent with observational results. On the other hand, \citet{DSouza2021} analysed the ELVIS suite of DM-only zoom-in simulations of MW-mass haloes and showed that mergers involving massive progenitors, i.e. massive accretions, do not lead to an increase in the total number of satellite subhaloes accreted by a host halo, but rather that they serve to cluster the satellites by their infall times. Such results indicate that satellite populations may contain signatures of the host's formation and merger histories, although it is still unclear which satellite properties best encode these signatures. It should be noted that although \citet{DSouza2021} considered DM subhaloes with peak total mass $\MGEN{peak}>10^{9}\MSUN$ which are assumed to host dwarf galaxies, it remains to be seen whether luminous satellites would show similar behaviour, and whether there are other, baryonic, properties that may show different correlations with merger histories.

Exploring the impact of merger histories on the satellite populations of galaxies would typically require either a cosmological simulation of a large enough volume to encompass a sample of merger histories representative of the Universe, or a suite of zoom-in simulations covering a range of merger histories. Furthermore, the simulations must have high-enough resolution to accurately model the dwarf population while simultaneously modelling the more massive host system. All of these factors incur significant computational costs. Additionally, in either method, it is not possible to control for other factors that would affect the satellite population along with the merger history itself. The \textsc{vintergatan-gm} suite of simulations \citep{Rey2022,Rey2023} attempts to circumvent the latter of these limitations by using the \textsc{GenetIC} algorithm \citep{Roth2016,Rey2018,Stopyra2021} to perform targeted modifications to the initial conditions (ICs) evolved by the simulation. The modifications change the mass ratio of a specified merger, while preserving the $z=0$ halo mass of the system and, to the greatest extent possible, the cosmological environment and surrounding structures. This allows us to isolate the impact of a single merger, while controlling for most other factors that would affect the evolution of the central galaxy and its satellites. As shown in \citet{Rey2023}, the modifications drastically alter the properties of the central galaxy, but its surrounding outer stellar halo is largely unaffected except in the case where the GMs result in a bulge-dominated central galaxy.

In this work, we focus on the satellite populations around the central galaxy at various times to understand their response to the central's merger history. We present a brief description of the simulations and methods in Section~\ref{sec:methods} and our results in Section~\ref{sec:results}. In Section~\ref{sec:discussion}, we discuss the physical mechanisms by which the merger histories affect the satellite populations and what this implies for techniques that make use of satellites to infer properties such as the host mass as well as its merger history. Our conclusions are summarized in Section~\ref{sec:conclusions}.

\section{Methods}   \label{sec:methods}

\subsection{Simulations}
This paper analyses the \textsc{vintergatan-gm} suite of simulations, which is comprised of five zoom-in cosmological hydrodynamical simulations, performed with the code \textsc{ramses} \citep{Teyssier2002}. The fiducial simulation of the suite focuses on a MW-mass system, whereas the other four simulations are variations where the ICs were genetically modified (GM) to alter the mass ratio of an important merger that occurs at $z \approx 2$, while maintaining the $z=0$ halo mass of the system, through the use of the \textsc{GenetIC} code \citep{Roth2016,Rey2018,Stopyra2021}. This suite of genetically modified ICs was first introduced in \citet{Rey2022}, where the authors performed multiple DM-only (DMO) zoom-in simulations of two MW-mass hosts. The target halo for the fiducial \textsc{vintergatan-gm} simulation was selected from these DMO zooms, themselves based on an initial uniform resolution DMO simulation of a $(50\ \text{Mpc})^3$ volume resolved by $512^3$ particles, with a DM particle mass resolution of $1.2 \times 10^{8}\MSUN$. The halo was chosen to be in the MW mass range of $\MGEN{200c} \approx 10^{12}\MSUN$, with no other massive neighbours within $5\RHOST$. Additionally, the halo was selected to have a significant merger at $z\sim 2$ and a relatively quiet merger history at later times, to approximate the inferred merger history of the MW. Here and throughout the paper, a significant merger is defined as one with i)~a mass ratio in $\MGEN{200c}$ more significant than 1:30 and b)~bringing in at least $\MSTAR \geq 10^{8}\MSUN$ at infall. These criteria identify any important mergers while excluding mergers at very early epochs which may have high mass ratios but are of relatively low importance in terms of mass content. We reserve the term `major' merger for only those with (DM) halo mass ratios $>1:3$ as is common practice in the literature. The $z\sim 2$ merger (rather than the most massive or most recent major merger) was targeted for the GMs to focus on the formation of the Gaia-Enceladus Sausage (GES), which is thought to have been accreted through such a merger \citep[see][for more details]{Rey2022,Rey2023}. The GM simulations alter the mass ratio of the target merger by altering the overdensity field at the position of the halo to 90, 95, 110 and 120 per cent of the fiducial value. Throughout the paper, we use the terms smaller/larger merger scenarios to refer to this decrease/increase in the $z \approx 2$ merger mass ratio and the labels Smallest/Smaller/Larger/Largest Merger B to refer to the corresponding simulations (the label `Merger B' is explained below in Section \ref{sec:results}.

Each of the simulations have a mass resolution of $m_{\text{DM}}=2.0\times 10^{5}\MSUN$ and an initial minimum gas cell mass of $m_{\mathrm{gas}}=3.6\times 10^{4}\MSUN$. The precise setup used for the simulations is described in detail in \citet{Rey2023}; we provide a brief description here. \textsc{ramses} is an Adaptive Mesh Refinement (AMR) based code which uses a particle-mesh algorithm to solve Poisson's equation and an HLLC Riemann solver for fluid dynamics assuming an ideal gas equation of state with $\gamma=5/3$. The AMR strategy allows us to reach spatial resolutions of 20 pc throughout the ISM. We employ the galaxy formation model of the \textsc{vintergatan} simulations \citep{Agertz2021}, which includes prescriptions for star-formation, feedback from SNeIa and SNeII, and stellar winds from O, B, and AGB stars. Stars are formed from cold dense gas ($\rho>100\ \rm{cm}^3$ , $T<100$ K) generating stellar particles with an initial mass of $10^{4} \MSUN$, modelled with a \citet{Chabrier2003} initial mass function. Feedback is injected in the form of thermal energy when the cooling radius is resolved by at least 6 gas cells, and in the form of momentum otherwise \citep{Kim2015,Martizzi2015,Agertz2021}. All simulations assume a flat $\Lambda$~CDM cosmology with $h=0.6727$, $\Omega_{m,0}=0.3139$, $\Omega_{b,0}=0.04916$, $\sigma_{8}=0.8440$ and $n_{s}=0.9645$ \citep{Planck2016} and linearly span cosmic time between $z=99$ and $z=0$.

\subsection{Halo finding and satellite selection} \label{sec:selection}
Haloes and subhaloes are identified within the simulation using the \textsc{ahf} halo finder \citep{Gill2004,Knollmann2009}, and only (sub)haloes consisting of at least 100 particles (of any type) are retained. Merger trees are constructed using the \textsc{pynbody} \citep{Pontzen2013} and \textsc{tangos} \citep{Pontzen2018} packages. Unless otherwise specified, (sub)halo properties are measured using all particles within the `halo radius', $r_{\text{halo}}$. \textsc{ahf} determines an initial $r_{\text{halo}}$ which is defined as either $\RHOST$ (in the case of haloes) or the subhalo-centric distance to the local minimum in the density field. It then iteratively removes unbound particles and defines a new radius enclosing the bound particles at each iteration, resulting in the final $r_{\text{halo}}$ provided in the halo catalogues.

From the halo catalogues, we select satellites around the central galaxy with the following criteria:
\begin{itemize}
    \item The stellar mass must satisfy $\MSTAR > 10^{6}\MSUN$ (i.e. resolved by more than $\sim100$ stellar particles), unless otherwise specified.
    \item The satellite should be found at a distance of $d/\RHOST > 0.15$ and $d/\RHOST <1$ unless specified otherwise. The central region is avoided to remove any unphysical clumps of stellar/gas matter belonging to the central galaxy from the final list of satellites.
    \item The satellite should have a baryon fraction $f_{\rm bar}=(\MGEN{*}+\MGEN{gas})/\MGEN{halo}<0.8$. This further removes any unphysical haloes with little or no DM, while still allowing for satellites that may have experienced significant tidal mass loss, which is likely to preferentially remove DM from the outer regions of the galaxy.
    \item We ignore subsubhaloes i.e. only retain haloes and their subhaloes. This ensures that we do not include small stellar/gas clumps within galaxies.
\end{itemize}

\section{Results}   \label{sec:results}

\begin{table*}
    \caption{Properties of the central MW-analogue galaxies in each of the GM simulations at $z=0$, start and end times ($t_{\rm start}$, $t_{\rm end}$) and merger mass ratios ($\mathcal{R_{\rm halo}}$ based on total halo mass, $\mathcal{R_{\rm stellar}}$ based on stellar mass) for all three significant mergers experienced by the central galaxy. All times are provided as cosmic age in Gyr. Merger B is the one targeted by the GMs, while mergers A and C are the preceding and following mergers. The start time is defined as the last time that the merging galaxy is outside the virial radius of the primary galaxy, whereas the end is defined as the time at which the merging galaxy coalesces with the primary. The merger mass ratios are measured using the host and secondary mass at the time of infall i.e. at $t_{\rm start}$ for each merger.} \label{tab:centralProps&MergerTimes}
    \centering
    \begin{tabular}{l|c|c|c|c|c|c|c|c|c|c|c|c|c|c|c}
    \hline
    Simulation & $\MGEN{200c}$ & $\RGEN{200c}$ & $\MSTAR$ & \multicolumn{4}{c}{Merger A} & \multicolumn{4}{c}{Merger B} & \multicolumn{4}{c}{Merger C} \\
    & [$10^{10}\MSUN$] & [kpc] & [$10^{10}\MSUN$] & $t_{\rm start}$ & $t_{\rm end}$ & $\mathcal{R}_{\rm halo}$ & $\mathcal{R}_{\rm stellar}$ & $t_{\rm start}$ & $t_{\rm end}$  & $\mathcal{R}_{\rm halo}$ & $\mathcal{R}_{\rm stellar}$ & $t_{\rm start}$ & $t_{\rm end}$ & $\mathcal{R}_{\rm halo}$ & $\mathcal{R}_{\rm stellar}$ \\
    \hline
    halo685x09 & 82.4 & 198.5 & 1.78 & 2.07 & 3.03 & 1:0.7 & 1:1.8 & 3.39 & 4.54 & 1:10.0 & 1:24.0 & 3.39 & 12.62 & 1:15.0 & 1:31.0 \\
halo685x095 & 81.3 & 197.7 & 1.55 & 2.02 & 2.86 & 1:0.9 & 1:2.1 & 3.24 & 4.41 & 1:9.8 & 1:15.0 & 3.53 & 13.28 & 1:14.0 & 1:20.0 \\
halo685 & 86.5 & 201.8 & 1.82 & 1.90 & 2.81 & 1:1.3 & 1:2.4 & 3.30 & 4.23 & 1:6.0 & 1:8.1 & 3.77 & 11.65 & 1:18.0 & 1:29.0 \\
halo685x11 & 84.4 & 200.2 & 1.61 & 1.90 & 2.81 & 1:1.4 & 1:1.8 & 3.30 & 4.36 & 1:2.9 & 1:4.3 & 3.83 & 11.10 & 1:19.0 & 1:26.0 \\
halo685x12 & 87.5 & 202.6 & 1.84 & 1.81 & 2.76 & 1:1.9 & 1:2.4 & 2.94 & 3.94 & 1:2.1 & 1:2.1 & 4.00 & 11.32 & 1:22.0 & 1:50.0 \\
    \hline
    \end{tabular}
\end{table*}

\begin{figure}
    \centering
    \includegraphics[width=\linewidth]{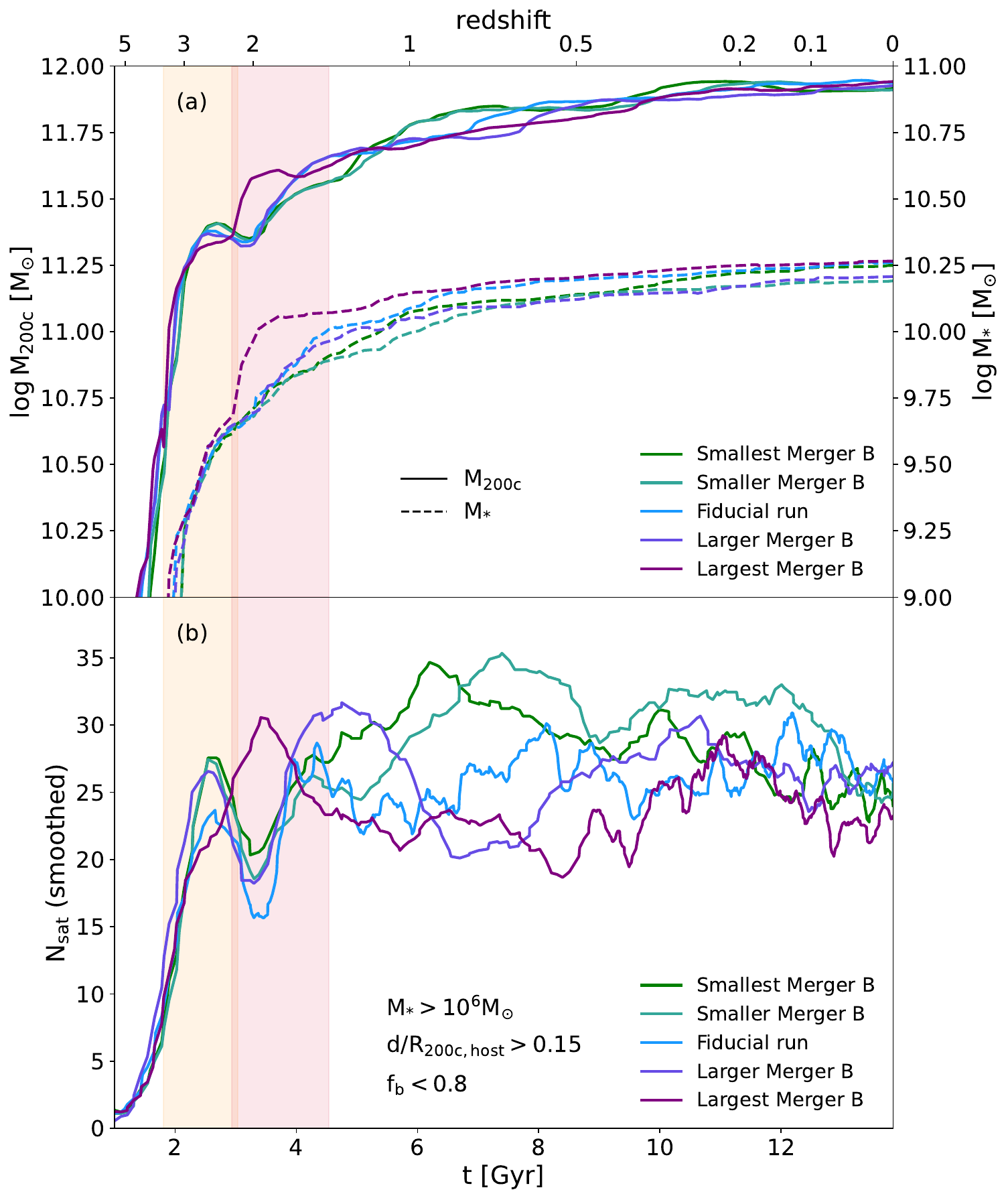}
    \caption{Evolution of the \emph{(a)} host halo mass and central stellar mass, and \emph{(b)} the number of satellites within the virial radius as a function of cosmic time. The number of satellites are a rolling average over 9 consecutive snapshots to smooth noisy data. The red shaded region indicates when the target merger (merger B) occurs while the orange shaded region indicates the earlier major merger A. Note that the precise start and end times for each merger are different between the five simulations (see Table \ref{tab:centralProps&MergerTimes}). The central galaxies have similar halo masses (by design), stellar masses and numbers of satellites at $z=0$. However, varying the mass ratio of merger B results in significantly different numbers of satellites at early times, particularly $\sim2-5$~Gyr after the end of merger B.} \label{fig:centralEvolMhaloMstarNsat}
\end{figure}

\begin{figure}
    \centering
    \includegraphics[width=\linewidth]{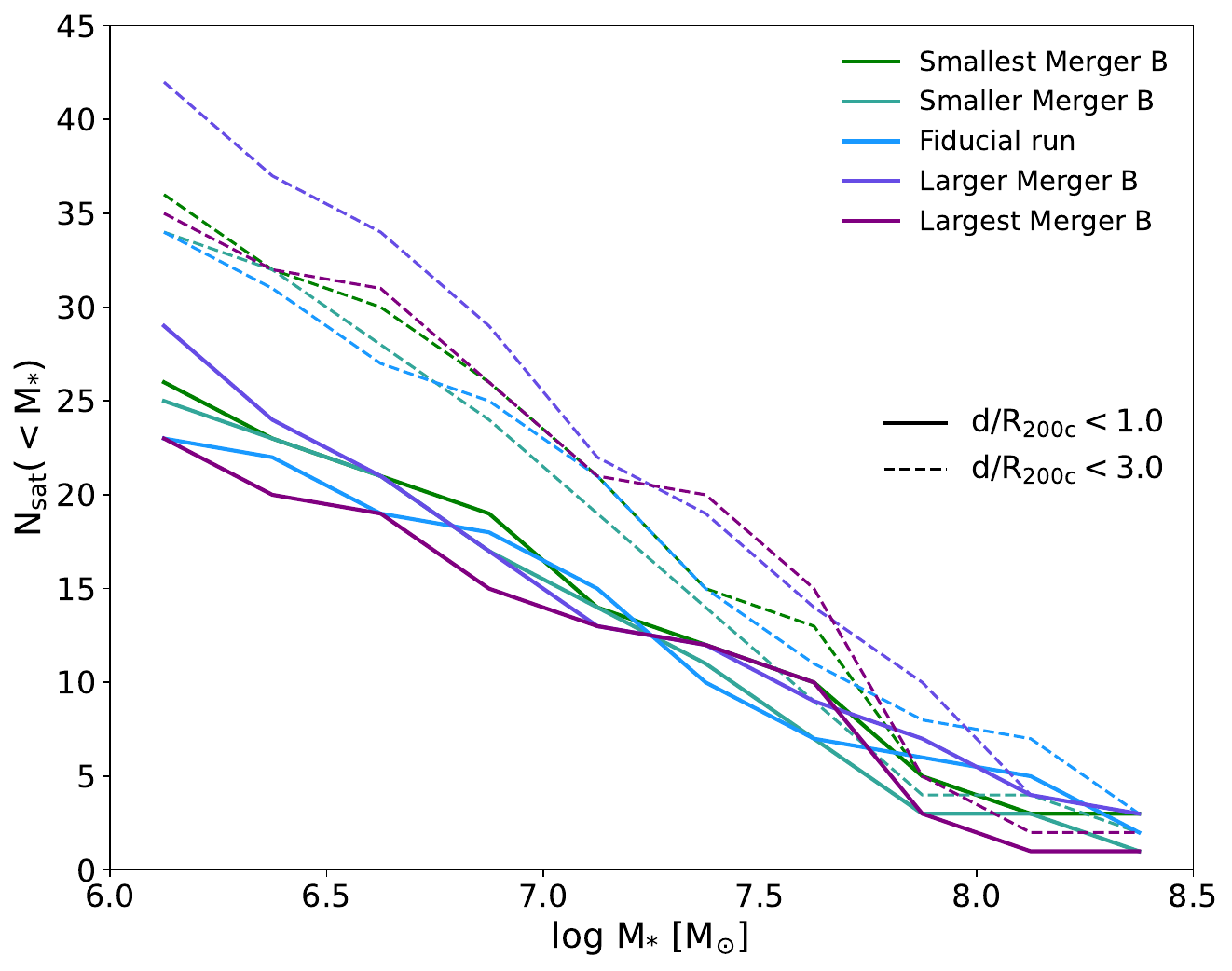}
    \caption{Stellar mass functions of satellites around the central galaxies for each of the GM simulations at $z=0$. Solid lines show satellites within $\RHOST$, dashed lines within $3\RHOST$. Note that these satellite numbers have not been corrected for observational biases or completeness effects. At $z=0$, there is no significant impact of the central's merger history on its satellite MF.} \label{fig:satMF}
\end{figure}

The genetic modifications applied to the initial conditions were designed to alter the mass ratio of the $z \approx 2$ merger experienced by the central galaxy in the fiducial simulation. Panel (a) of Fig.~\ref{fig:centralEvolMhaloMstarNsat} shows the resultant evolution of the central galaxies' halo (solid curves) and stellar (dashed curves) masses over cosmic time for each of the five GM simulations. Additionally, Table~\ref{tab:centralProps&MergerTimes} provides some key properties of the central galaxies at $z=0$. As seen from Fig.~\ref{fig:centralEvolMhaloMstarNsat}(a) and Table~\ref{tab:centralProps&MergerTimes}, by design, the final halo mass only varies by a maximum of $6\%$, and the virial radius by at most $2\%$, compared to the fiducial simulation. The stellar mass can vary by up to $14\%$, showing that modifications to early merger histories result in only a small amount of scatter in the stellar mass-halo mass relation. Due to the correlations implicit in a $\Lambda$CDM cosmology, genetic modifications by construction alter not only the mass ratio of the $z \sim 2$ merger (hereafter referred to as merger B) but also its timing, as well as the properties of two other significant mergers, one each before and after the target merger \citep[hereafter referred to as mergers A and C respectively; see][for complete details]{Rey2023}. The three mergers were identified in the fiducial simulation based on the criteria for significant mergers described in Section \ref{sec:methods} i.e. mass ratio $>1:30$ and $\MGEN{*,at\ infall} \geq 10^{8}\MSUN$. The merging systems are then cross-matched in the GM simulations. The start and end times of each of the three mergers are provided in Table \ref{tab:centralProps&MergerTimes} along with the corresponding merger mass ratios based on both total halo mass and stellar mass. The start time is defined as the final time that the merging galaxy is outside the virial radius of the primary galaxy, i.e. its infall time, whereas the end is defined as the time of coalescence. The merger mass ratios were measured at the time of infall of the merging galaxy as detailed in \citet{Rey2023}. Overall, our results reflect the combined impact of (i) modifying the target merger (ii) the necessary compensations in other mergers to reach the final halo mass within a $\Lambda$CDM cosmology and (iii) the non-linear interactions between these resulting from gravitational and baryonic evolution. Such interactions and cosmology-induced correlations are to be expected across all populations of observed or simulated galaxies, and must be taken into account when interpreting observational constraints on merger histories \citep{Rey2023}.

\subsection{Satellite mass functions at $z=0$}

We first compare the satellite mass functions around each of the central galaxies at $z=0$ in Fig.~\ref{fig:satMF}. Solid lines show the satellites within $\RHOST$ and dashed lines within $3\RHOST$. We include the latter since the effects of the merger may extend beyond the virial radius itself. The host merger history does not have a significant correlation with either the shape or the normalization of the satellite MFs within either spatial extent. Several previous studies have examined the correlation between satellite abundances and host halo ages/assembly times and have found that hosts that formed at later times on average contained more satellites \citep[e.g.][]{Gao2004,Mao2015,Artale2018,Zehavi2018,Bose2019}, suggesting that in earlier forming haloes, satellites had more time to merge with the centrals. Although our results seem at odd with those results, note that these previous studies considered satellites that are significantly more massive that those in our sample ($\MSTAR \gtrsim 10^{9}\MSUN$) and do not consider the number of satellites per halo, but rather average satellite populations at a given halo mass. Furthermore, as we show in later sections, we do find similar results when considering the mass functions at earlier times. On the other hand, studies such as \citet{Bose2020} have found the opposite trend i.e. that later forming haloes had fewer ultrafaint satellites ($\MSTAR \lesssim 10^{5}\MSUN$), when including orphaned galaxies (i.e. DM haloes that have been disrupted below the detection limit in DM-only simulations, but that are tracked beyond this point based on their orbital properties). These results highlight that the connection between halo assembly and satellite abundance is not fully established and that satellite masses are key when making such comparisons.

While comparisons between simulations may be performed in this way, comparisons to observations such as from the SAGA survey will require careful consideration of a number of different factors including: (i) surface brightness limits affecting both the sample selection and the measured masses/luminosities, (ii) the precise selection criteria used to define satellites, which may be different in simulations and observations, and (iii) the impact of cosmic variance, i.e. variations from galaxy to galaxy based on their local environment and aspects of their history beyond those systematically varied in our study. This variance will be especially important at the high mass end, where numbers are small.

We will tackle a comparison to observations in future work, but preliminary analyses indicate that a significant fraction of the low-mass satellites included in our current sample have surface brightness values too low for them to be detected in observations and would therefore not be included in observed samples. The results are also  dependent on the precise aperture and filter used to measure mock luminosities. As such, we defer conclusions about direct comparisons between our simulations and observations to a future paper.

\subsection{The evolution of satellite populations} \label{sec:satEvolution}

\begin{figure}
    \centering
    \includegraphics[width=\linewidth]{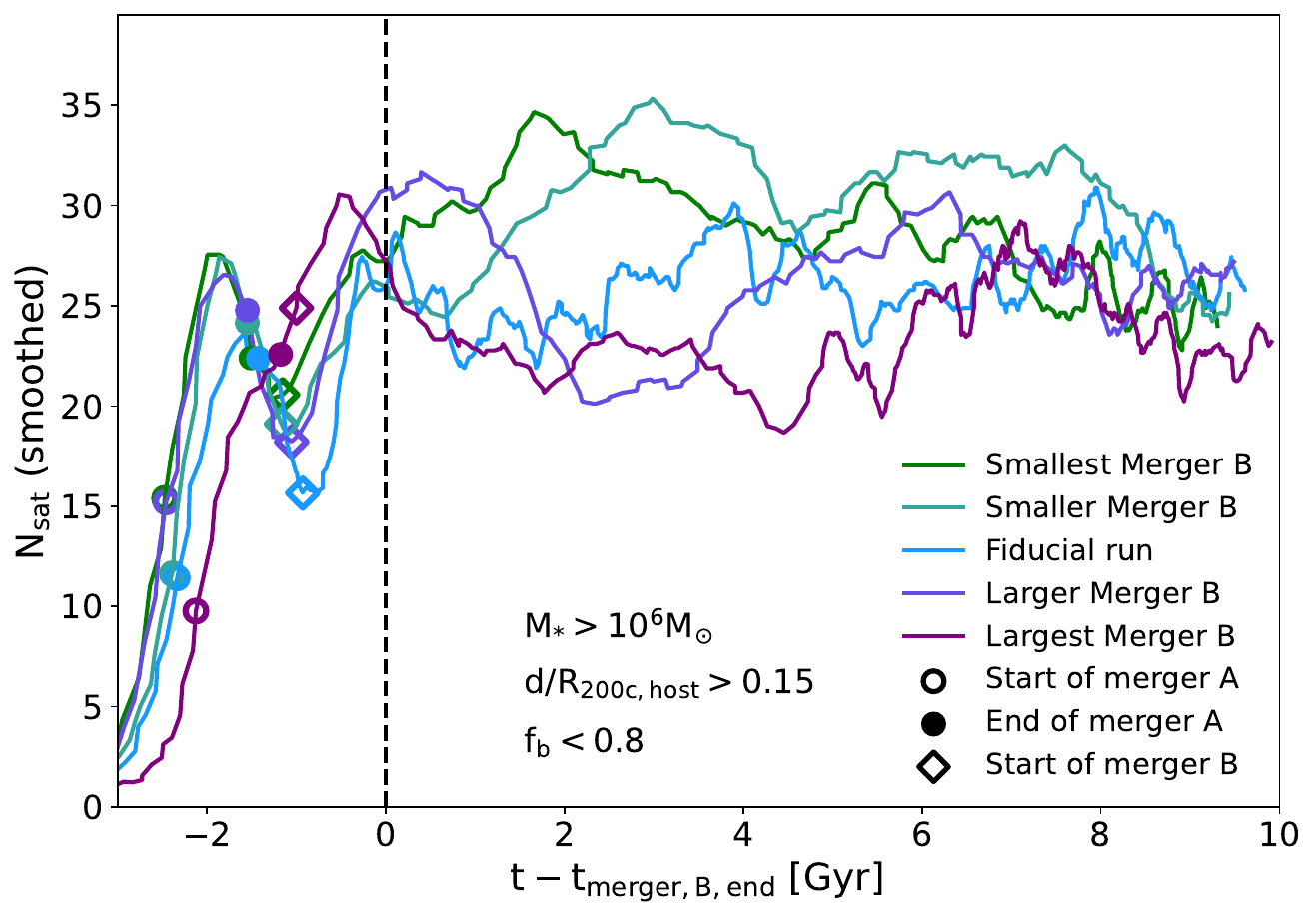} \caption{Evolution of the number of satellites around the central galaxy, as with Fig. \ref{fig:centralEvolMhaloMstarNsat}(b), but as a function of time relative to the end of merger B. Open and filled circle markers indicate the start and end of merger A respectively, while diamond markers indicate the start of merger B. The first half of merger A is characterized by a rapid increase in satellite numbers; the second half of merger A shows a steady decline in satellite numbers until the beginning of merger B, except in the case of the \XCXX{} simulation, where no such decline is seen.} \label{fig:nSatEvolutionWRTMerger}
\end{figure}

\begin{figure}
    \centering
    \includegraphics[width=\linewidth]{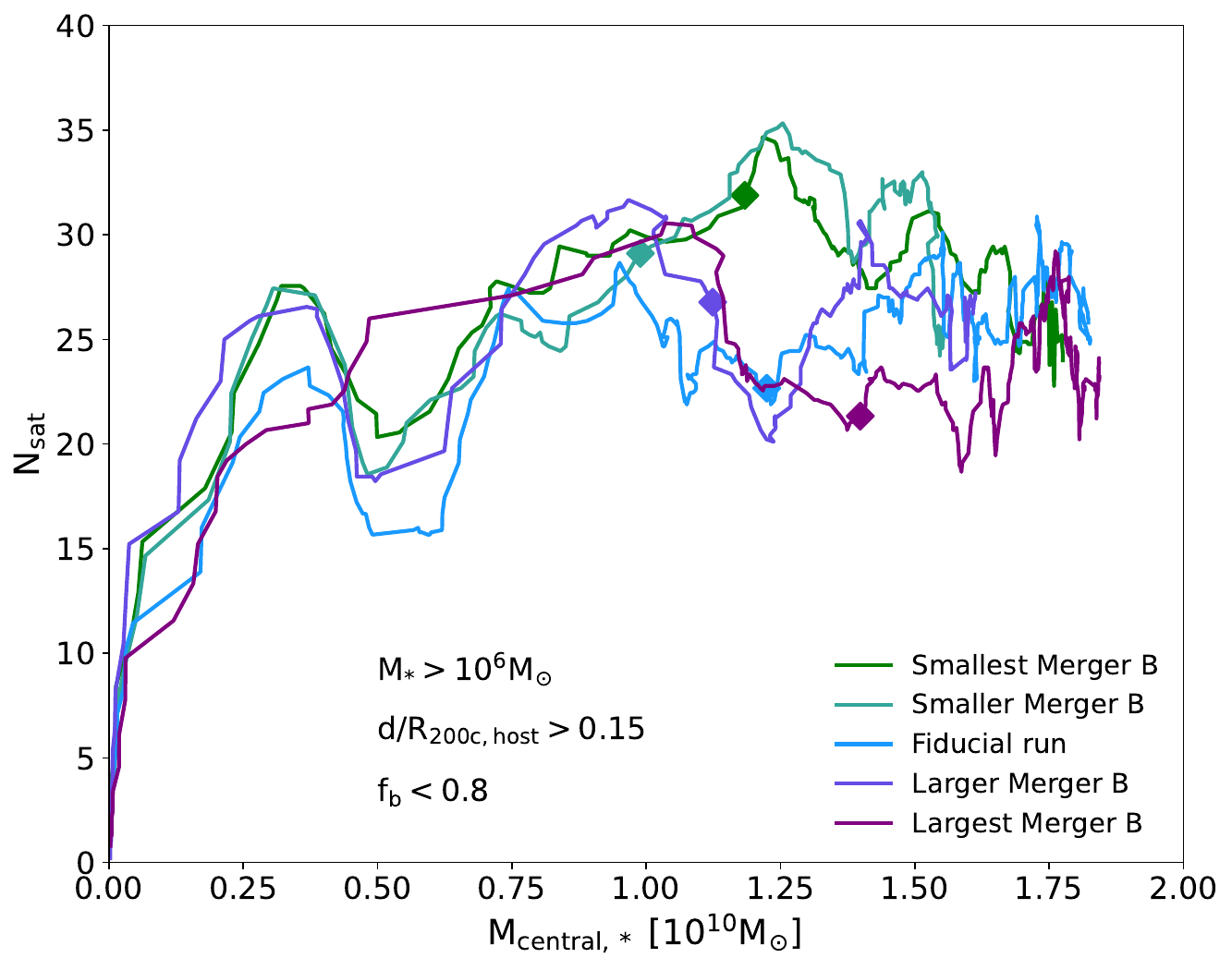}
    \caption{Evolution of the number of satellites within the virial radius around the central galaxy, as a function of the  central stellar mass. The diamond markers indicate the location of the central galaxies on this plane at $z \approx 1$. Note that the raw data are noisy and have therefore been smoothed using a running average over 9 consecutive snapshots. The number of satellites can vary by as much as 10-15 at a fixed stellar mass, indicating that the difference in satellite numbers seen in previous figures are not simply due to different mass growth histories of the central galaxies.} \label{fig:nSatEvolutionMStar}
\end{figure}

\begin{figure*}
    \centering
    \includegraphics[width=\linewidth]{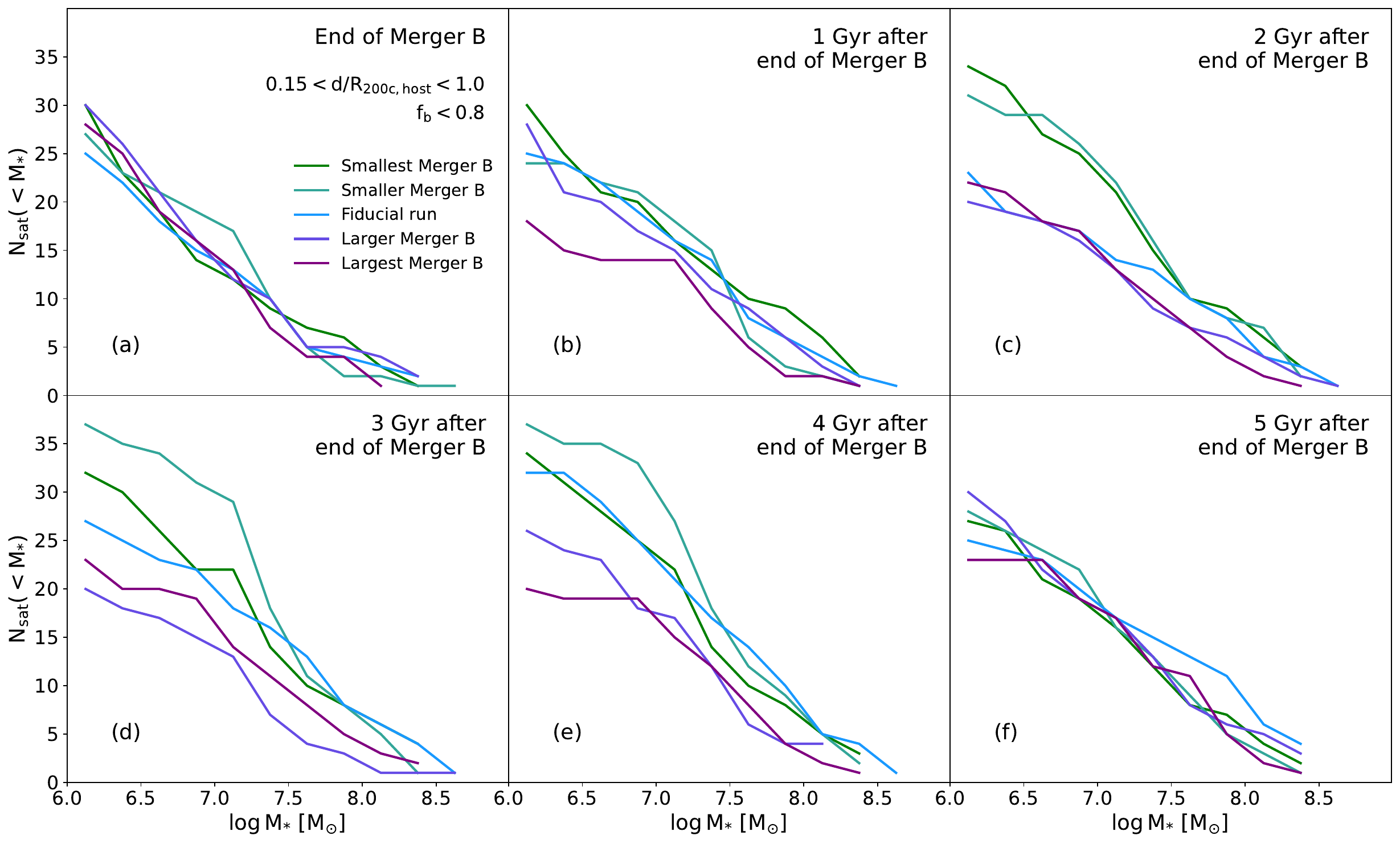}
    \caption{Satellite MFs at the end of the GM target merger (merger B) and up to 5 Gyr after the end of merger B for each of the GM simulations. Starting with similar mass functions of satellites at the end of merger B, the smaller merger scenarios result in an increase of low-mass ($\MSTAR \lesssim 10^{7.5}\MSUN$) satellites. The impact of the merger remains visible for $\sim2-5$ Gyr after the end of the merger, with the MFs becoming indistinguishable from one another after 5 Gyr.} \label{fig:satMFAfterMerger}
\end{figure*}

We next turn our attention to the time evolution of the satellite MFs. Fig. \ref{fig:centralEvolMhaloMstarNsat}(b) shows the evolution of the total number of satellites around the central galaxy as a function of cosmic time. Note that the raw data are noisy, partly due to some subhaloes not being detected by \textsc{ahf} especially during mergers or in very high-density regions, and have therefore been smoothed with a rolling average. While the number of satellites is similar at $z=0$ for each of the GM simulations, this is not the case at earlier times. At the beginning of merger A (orange shaded region), each of the GM simulations have approximately the same numbers of satellites. The start time of this merger varies by at most $\sim 100$ Myr across the GM simulations (see Table~\ref{tab:centralProps&MergerTimes}). The start time of merger B is more variable, varying by up to $130$ Myr across four simulations and in one exceptional case, the \XCXX{} simulation, occurring $350$ Myr earlier than in the fiducial case. Although it appears that the end of merger A overlaps with the beginning of merger B in Fig. \ref{fig:centralEvolMhaloMstarNsat}, this is merely the result of combining the start and end times of all five simulations and in fact, within each simulation, there is a time interval of $\sim 200-500$ Myr between the two events. 

The combined impact of these differences in the merger timing and mass ratios is evident in Fig.~\ref{fig:centralEvolMhaloMstarNsat}(a), where the largest merger scenario exhibits an earlier build up of halo and stellar mass at $z \sim 2$ compared to the other simulations, and in Fig.~\ref{fig:centralEvolMhaloMstarNsat}(b), where the evolution of the number of satellites is markedly different, especially at $z \lesssim 1.5$.

To show the effects of merger B more clearly, in Fig. \ref{fig:nSatEvolutionWRTMerger}, we show the same evolution of number of satellites, but now as a function of time relative to the end of merger B (i.e. the time of coalescence of the merging system). The start and end of merger A are indicated as open and filled circle markers, and the start of merger B as diamond markers. During merger A, the central halo rapidly accumulates several satellites. Despite some differences in timing, all five simulations have similar numbers of satellites by the end of merger A ($\Delta N (t_{\text{end,mergerA}}) \sim 2$).

In all cases except for the \XCXX{} simulation, the number of satellites then shows a sharp decline, starting shortly before the end of merger A and continuing until the beginning of merger B.  During merger B, the hosts further accumulate additional satellites, such that by the end of merger B, they again have similar numbers of satellites ($\Delta N (t_{\text{end,mergerB}}) \sim 7$). Roughly $2-3$ Gyr later however, there is a stark difference in the number of satellites ($\Delta N (t_{\text{end,mergerB}}) \sim 11-12$), with the smaller merger scenarios resulting in a higher number of satellites. In fact, while the smaller merger scenarios continue to accumulate satellites, the number of satellites is seen to decrease in the case of the fiducial simulation and larger merger scenarios. We discuss the likely causes for these trends in Sec. \ref{sec:tracking}.

In order to determine to what extent the previous results are due to differences in the mass of the central at a given time, in Fig. \ref{fig:nSatEvolutionMStar}, we show the evolution of the number of satellites as a function of the central stellar mass. Diamond markers indicate the mass and number of satellites at $z \sim 1$ for comparison. The figure shows that even after controlling for the mass of the central, there are significant differences in the numbers of satellites, with the smallest merger scenario having $\sim 10$ more satellites than the largest merger scenario, although this trend is not monotonic with the mass ratio of merger B. We have confirmed that the results are similar when considering the hosts' halo masses instead of stellar mass. The trends are somewhat weaker when considering satellites out to $3\RHOST$, but qualitatively similar, indicating that the scatter is not due to satellites simply travelling beyond the virial radius. Hence, the differences in numbers of satellites soon after the merger persist even when controlling for the central mass, which implies that the central merger history may be a significant source of uncertainty when recovering the properties of a group/cluster, e.g. host mass, from its satellite populations. 

These results are broadly consistent with recent observational results from \citet{Wu2022}. While that work found a strong dependence of number of satellites on host stellar mass, their results span a large range of central galaxy stellar masses compared to our sample and averages over several host-satellite systems. From Fig. \ref{fig:nSatEvolutionMStar}, we can approximate the maximum scatter in $N_{\rm sat}$ at a given host stellar mass to be $\sim 35-40$\% for masses of $\MGEN{central,*} \sim (1-1.75)\times 10^{10}\,\MSUN$, whereas \citet{Wu2022} find a scatter of $\sim 20-25$\% for satellites within 200-300 kpc from the host for host central stellar masses of $10^{10-10.25}\MSUN$ (see fig. 5 in their paper). Note that the observed satellites are significantly more massive ($M_{r}<-15.0$) compared to the satellites under consideration here, which likely accounts for the lower scatter in the observational results.

\citet{Smercina2022} observationally infer the mass of mergers using stellar halos, and find a strong correlation between satellite richness and mass of the most dominant merger secondary. However they must infer the mass of the merger indirectly from the stellar halo, and furthermore have a sample which covers a wide range of central masses, unlike our controlled simulations. Our result that mergers have a modest long-term effect on the satellite population are thus not in direct tension with the strong correlation reported by \citet{Smercina2022}, and further future work would be required to provide a direct point of comparison. Our results are also broadly in agreement with those of \citet{DSouza2021}, exhibiting a similar rise in number of DM subhaloes during massive accretions, but soon after followed by a rapid decrease in satellite numbers. The luminous satellites in our simulations also show similar a similar rise and rapid drop in satellite numbers (Fig. \ref{fig:centralEvolMhaloMstarNsat}(c) \& \ref{fig:nSatEvolutionWRTMerger}).

\subsection{Variation of satellite MFs over time}

Fig. \ref{fig:nSatEvolutionWRTMerger} shows that there are significant differences in the number of satellites that persist for several Gyr. We now examine the satellite MFs over this time period to understand the mass dependence of the previous results. Fig. \ref{fig:satMFAfterMerger} shows the stellar MFs of satellites within $\RHOST$ at and 1-5 Gyr after the end of merger B. The MFs remain similar for the first Gyr in all cases except the \XCXX{} simulation, which shows a noticeable loss of low- and intermediate-mass ($\MSTAR \lesssim 10^{7.5}\MSUN$) satellites. By 2 Gyr after the merger however, we find significant differences between the smaller and larger merger scenarios, with the former (latter) having gained (lost) several low-mass satellites. These differences are seen to persist for approximately 3 Gyr; by $\sim 5$ Gyr after the end of merger B however, the satellite MFs in all the GM simulations are once again indistinguishable from one another. Thus, the differences in satellite numbers found in previous results are largely driven by the loss/gain of low-mass satellites, while the number of more massive satellites remain approximately constant, changing by at most a few over a $\sim 4-5$ Gyr interval after the end of a significant merger. To quantify more concretely the timescale over which the impact of the merger is noticeable, we compare the satellite cumulative MFs (cMFs) in 0.25 Gyr increments after the end of merger B and record the maximum difference between the \XC{} simulation and each of the other simulations in any mass bin, normalized by the $\sqrt{\rm cMF}$ of the \XC{} simulation. The merger impact timescale can then be measured as the time over which this maximum difference remains greater than $2\sqrt{\rm cMF}$, which ranges from $2.25-4.25$ Gyr across the five simulations. This can be compared to the dynamical time of the system at the virial radius at the end of merger B, which is $1.17-1.33$~Gyr across the five simulations; thus the satellite MFs respond to the target merger over $\sim2-4$ dynamical times as expected.

While increasing numbers of low-mass satellites can easily be tied to accretion of associated subhaloes, the mechanisms for declining numbers is harder to pinpoint and can include a) travelling beyond the virial radius of the host i.e. becoming \emph{backsplash} galaxies \citep{Gill2005}, b) merging with the central, or c) being disrupted below the detection limits of the halo finder or losing enough mass due to tidal stripping to fall below the stellar mass limit we have imposed. The first of these suggests that the splashback radius is a more physically motivated definition of the host halo boundary rather than the virial radius when determining satellite membership, as has been proposed by several previous studies \citep[e.g.][]{Adhikari2014,More2015,Diemer2017}. However we have conducted the same analysis including satellites out to 2 and $3\RHOST$ and while there are indeed differences in the numbers of satellites, especially between mergers A and B, neither choice changes the broad conclusions of this paper.

Ideally one would quantify the contribution of these different effects to shaping the satellite population by identifying the fate of each individual subhalo. However, the currently available merger trees are not robust enough to be able to track individual subhaloes through the mergers in order to determine precisely which of these pathways they follow. While \textsc{ahf} can detect most subhaloes around the central halo, during close pericentric passages, we find that subhaloes are often identified as merging into the central galaxy, partly due to physical processes but partly due to misidentification of subhalo particles. This is especially true during mergers, when there are two (and sometimes more) central galaxies, making it challenging to automatically track the satellite galaxies through the mergers to determine which of these pathways are most important. Instead, in Section \ref{sec:tracking}, we use a combination of manually inspection and tracing individual stellar particles to understand how the merger history shapes the satellite population.

\section{Discussion}  \label{sec:discussion}

\begin{figure*}
    \centering
    \includegraphics[width=\linewidth]{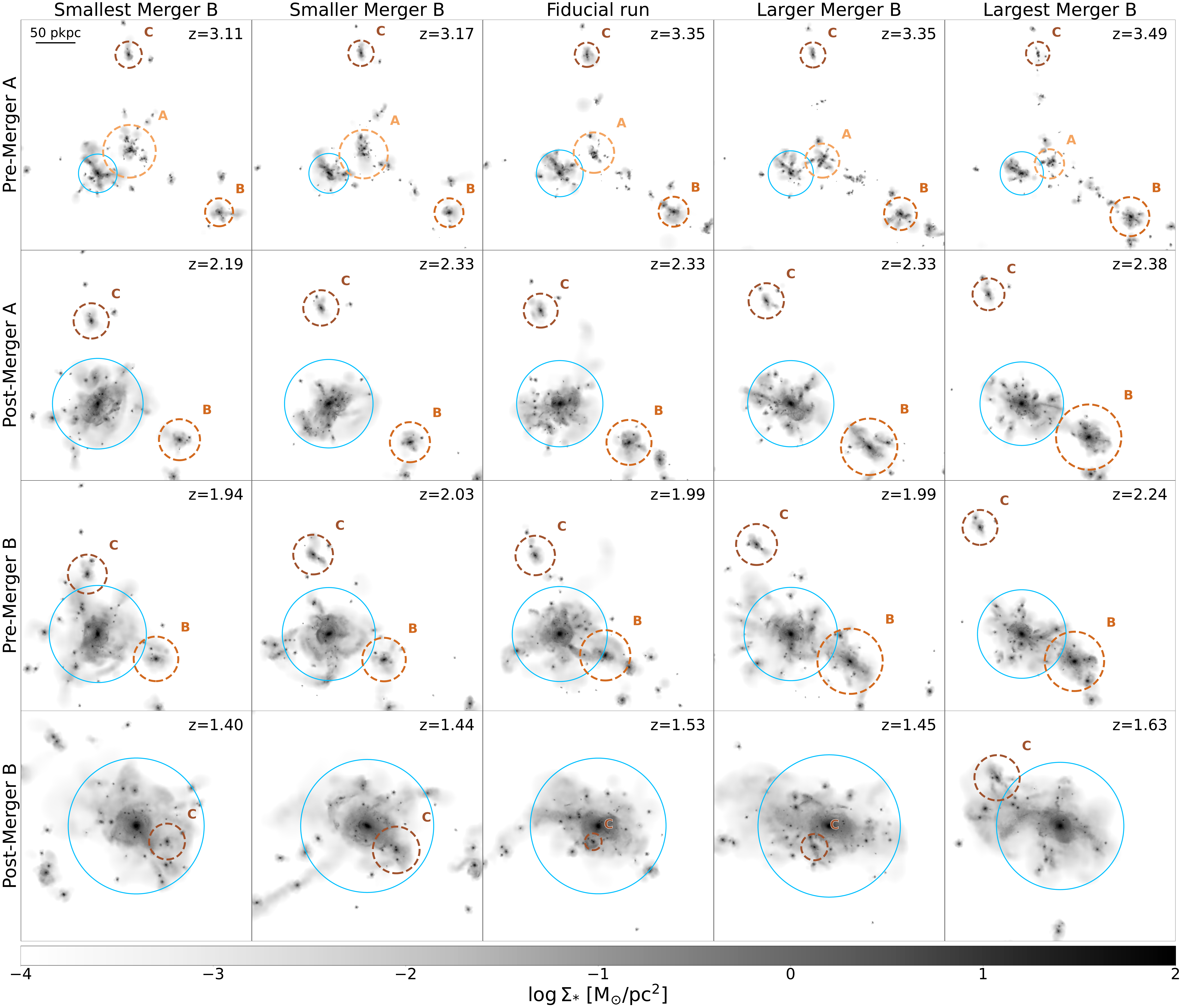}
    \caption{Projected stellar mass distribution at the beginning and end of mergers A (rows 1 \& 2) and B (rows 3 \& 4) in each of the GM simulations. The blue circle marks the virial radius of the main host, while the brown dashed circles (lightest to darkest) indicate the merging systems in mergers A, B and C respectively. Each image has a depth of $\pm 200$~kpc relative to the centre of the main halo; while each image is of size 300 kpc $\times$ 300 kpc, the images in the upper 3 rows are not centred on the main halo to show the other merging systems. Row 1 shows the impact of the GMs on the merging system A, which becomes progressively smaller and contains fewer satellites as merger B becomes more important. Furthermore, there are several small galaxies present in the region between the merging systems A and B, increasing in number from the \XXC{} to \XCXX{} scenarios, which contribute to the increase in satellite numbers seen at the end of merger A in Figs. \ref{fig:centralEvolMhaloMstarNsat}(c) and \ref{fig:nSatEvolutionWRTMerger}.}
    \label{fig:stellarMaps}
\end{figure*}

\begin{figure*}
    \centering
    \includegraphics[width=\linewidth]{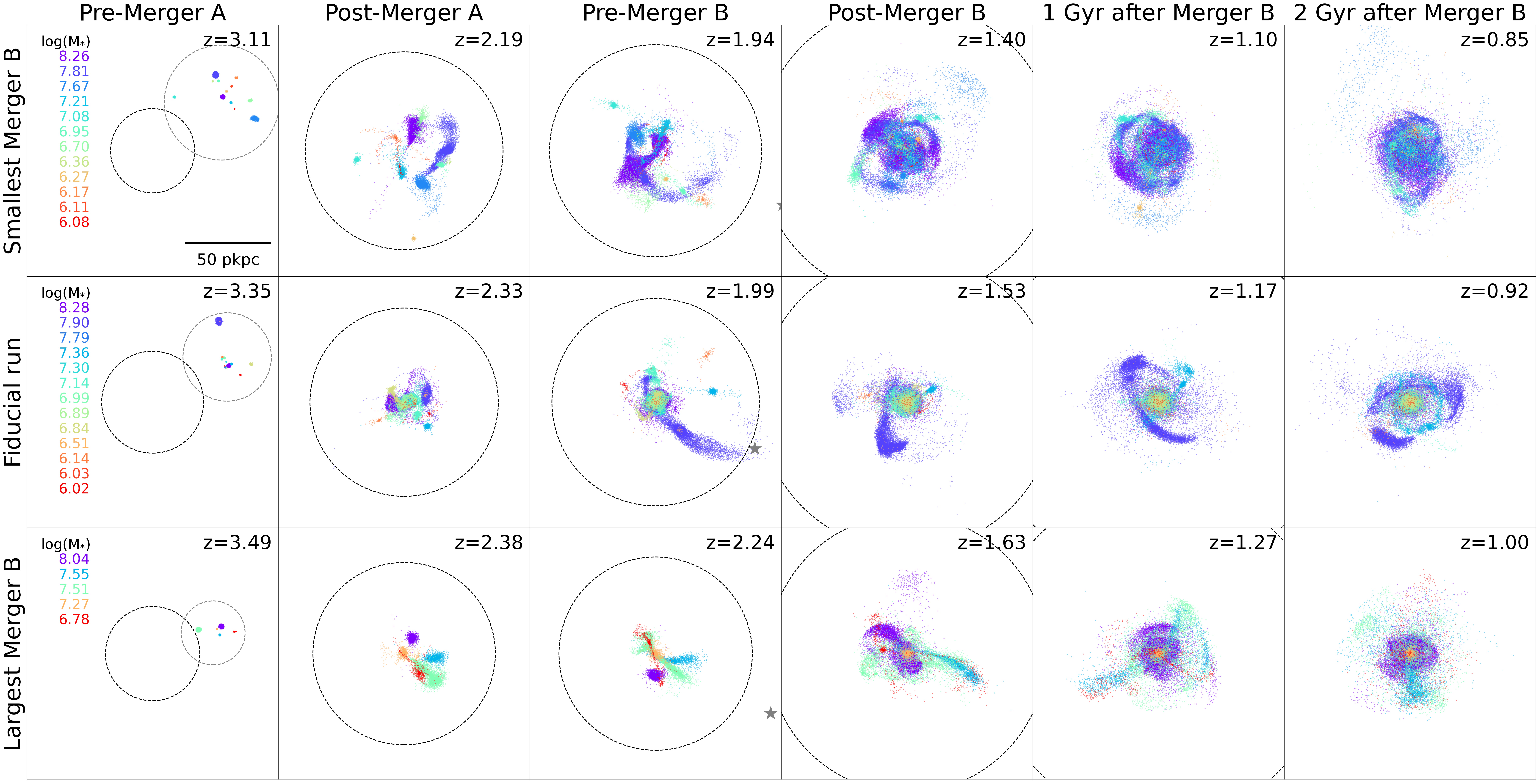}
    \caption{Projected map of stellar particles from satellites brought in by the secondary system in merger A for the fiducial run (row 2) and the smallest and largest merger scenarios (rows 1 and 3). Satellites are identified at the beginning of merger A (shown in column 1) and tracked in subsequent snapshots showing the end of merger A, beginning and end of merger B, and 1 \& 2 Gyr after the end of merger B (columns 2-6). Particles are coloured by the stellar mass rank of the satellites they originate from in the first snapshot (rather than stellar mass itself, to maximize contrast between satellites). For clarity, we select particles within $0.25 \times \RGEN{halo}$ for each satellite in the first snapshot. The black dashed circle shows the extent of the main halo, while the grey dashed circle shows the location of the merging system. Additionally, in the ``Pre-Merger B'' panels (column 3), we indicate the position of the merger B secondary progenitor with a grey star; at earlier times (columns 1 \& 2), this system is beyond the range of positions plotted. The majority of satellites brought in during merger A have been disrupted by the beginning of merger B.}
    \label{fig:satTrackingA}
\end{figure*}

\begin{figure*}
    \centering
    \includegraphics[width=\linewidth]{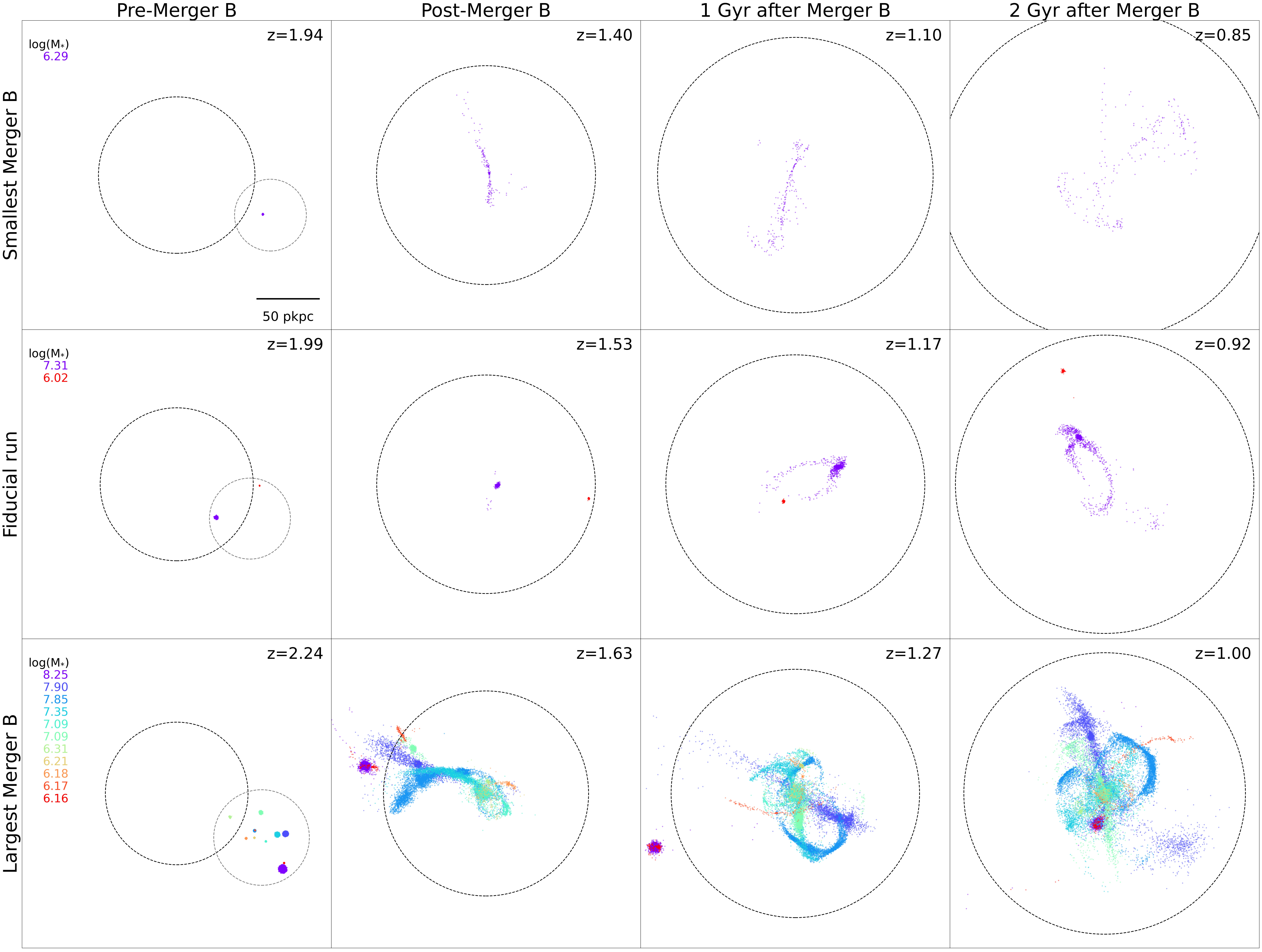}
    \caption{Same as Fig. \ref{fig:satTrackingA} (cols 3-6), but now showing satellites brought in through merger B with the merging system indicated by the grey dashed circle. Merger B introduces fewer new satellites to the system, but these satellites are longer lived. However, some can be seen to travel beyond the virial radius and also be partially disrupted, leading to a loss of stellar mass and thus making them less massive than required by our selection criteria.}
    \label{fig:satTrackingB}
\end{figure*}

\subsection{Satellite accretion histories} \label{sec:tracking}
The results described above show that altering the mass ratio of a merger experienced by a MW-mass galaxy also changes how it accretes its satellite population. The genetic modification approach highlights the non-linearities implicit in galaxy formation physics and the correlations required by a $\Lambda$CDM cosmology, implying that any single merger (whether observed or simulated) cannot be interpreted in isolation from other accretion events. While the target for modification is merger B, in order to achieve the same halo mass at $z=0$, the modifications must indirectly affect several other mergers, notably the merger that occurs immediately before (merger A). An increase/decrease in the mass ratio of merger B is partially compensated for by a decrease/increase in the mass ratio of merger A \citep[for details see][]{Rey2023}. As mentioned in Section \ref{sec:satEvolution}, the limitations of the halo finder and the merger trees prevent us from reliably tracking individual satellites to quantify the role of different physical effects, especially when considering how satellites are eventually destroyed. Here we discuss the various ways in which the genetic modifications affect the merging systems, which can qualitatively shed light on this issue.

We first look at the commonalities between the merger histories of the five simulations and then highlight the differences between them. Some of the key features of the merger histories are illustrated in Fig. \ref{fig:stellarMaps}, which shows the projected distribution of stellar particles within 200~kpc from the central halo for each of the GM simulations (columns) at the beginning and end of mergers A (rows 1 \& 2) and B (rows 3 \& 4). The main halo's virial radius is shown by the blue solid circle, while the merging systems A, B and C are shown in dashed brown circles (lightest to darkest). 

In general, the merger histories are composed of the following sequence of events:
\begin{enumerate}
    \item Prior to merger A (the first $\sim 2$~Gyr of the simulation), the galaxy's main progenitor and the secondary merging system both experience a chaotic fast accretion period undergoing several close interactions, often involving multiple simultaneously interacting galaxies. Some, if not most, of these interacting galaxies do not have enough time to merge with the corresponding central galaxy, instead becoming part of the satellite populations of the two merging central galaxies.
    \item Merger A proceeds, beginning with the secondary central galaxy crossing the virial radius of the primary system, followed by a first pericentric and then first apocentric passage. Eventually, the secondary central galaxy coalesces with the primary central after a few orbits, while the two satellite systems become mixed. Between the beginning of merger A and the first pericentre, the number of satellites increase, resulting in the first peak in satellite numbers at $z \sim 2.5$ in Fig. \ref{fig:centralEvolMhaloMstarNsat}(b).
    \item Following the first pericentric passage and before the beginning of merger B, while some satellites remain in the system, of the ones that do not, the satellites can follow one of three different pathways: (a) some have velocities large enough to travel beyond the virial radius of the merged system, i.e. become backsplash galaxies, (b) some are temporarily discounted since they are within $0.15 \RHOST$ (or indeed are not detected by the halo finder at all within the high density central region), or (c) some merge with the central galaxy. All three of these together result in a decrease in the number of satellites at $z \sim 1.9$ (this is discussed in more detail below.) Note that we find that satellites can have a wide range of merger timescales, with some satellites (usually low mass and/or on radial trajectories) merging within $<1$~Gyr, while some satellites (usually more massive and/or on circular trajectories) can remain for several Gyrs. This confirms that satellite merging cannot account for the entire decrease in satellite numbers at $z \sim 1.9$.
    \item Between mergers A and B, the system can also interact with several smaller infalling galaxies, which do not necessarily merge immediately, but instead add to the satellite population.
    \item Merger B proceeds, with the secondary system bringing in some satellites of its own, leading to an increase in the number of satellites beginning at $z \sim 1.9$. The satellites brought in by this second merger also follow multiple pathways as mentioned above for merger A, and discussed in more detail below.
    \item The post-merger-B phase begins, consisting of the third significant merger, C, and several other smaller mergers.
\end{enumerate}

Given this sequence of events which is in common between all simulations, we can now discuss in detail where the merger histories differ, and how this gives rise to differences in the satellite population. 
\begin{itemize}
    \item The secondary system involved in merger A assembles during step (i) above, and is most responsible for compensating for the increased/decreased mass budget of merger B in the GM simulations. Thus, it brings in more satellites in the \XXC{} simulation (12 with $\MSTAR>10^{6}\MSUN$) and fewer in the \XCXX{} simulation (5), which goes to explain the relative sizes of the peaks seen at $z \sim 2.5$ in Fig.~\ref{fig:centralEvolMhaloMstarNsat}(b). This is also evident in Fig. \ref{fig:stellarMaps} (top row), where the size of the merging system A gets progressively smaller as the importance of merger B increases (left to right). This reinforces how even the \textit{observed} effect of a significant merger within a $\Lambda$CDM cosmology will not be fully separable from the events leading up to that merger; an observed set of galaxies with different ongoing or recent mergers, even if controlled for fixed central galaxy mass, will exhibit both direct and indirect consequences of the ongoing merger.
    \item During step (iii), as the satellites brought in with merger A settle onto new orbits, two key factors differ between the scenarios:
    \begin{itemize}
        \item The numbers of satellites that are travelling beyond the virial radius decrease from the \XXC{} to \XC{} to \XCXX{} simulations, due to the increased kinetic energy of A and therefore its satellite system in the former cases.
        \item The interval between the first pericentric passage during merger A and the beginning of merger B, is shorter for the \XCXX{} scenario and longer for the \XXC{} one. In fact, merger B begins $400-450$ Myr earlier for the \XCXX{} simulation than for the other four simulations.
    \end{itemize}
    \item Between merger A and B, the system further accretes smaller galaxies: 3 in the \XXC{} simulation, 10 in the \XC{} simulation (very shortly after merger A itself) and 8 in the \XCXX{} simulation (approximately evenly spread out between mergers A and B). These can also be seen in the top row of Fig. \ref{fig:stellarMaps}, as several smaller galaxies spanning the region between systems A and B, with the fewest seen in the \XXC{} simulation and most in the \XCXX{} simulation. These are then directly responsible for the increase in satellite numbers seen shortly after the end of merger A, and in fact, in the case of the \XCXX{} simulation, for completely erasing the decrease expected after merger A. In effect, the satellites delivered with merger A in the \XXC{} simulation are instead delivered more gradually between mergers A and B in the \XCXX{} case.
    \item The secondary system in merger B contains at most 2 satellites with $\MSTAR>10^{6}\MSUN$ in the three smaller merger B simulations, but 5 and 11 in the two larger merger B simulations. From the \XXC{} scenario to the \XCXX{} one, the central mass of B grows significantly (by construction), and the number of small galaxies in its vicinity also grows as is evident from Fig. \ref{fig:stellarMaps} (rows 2 and 3). However, these small accreted systems are not formally considered satellites of B, and so can be seen as an extension of the enhanced satellite capture rate already described between the two mergers.
    \item Finally, after merger B, the system continues to accrete smaller galaxies. The number of such events is mildly smaller in the \XCXX{} scenario than in the \XXC{} one and they occur at slightly later times; more importantly, the galaxies are noticeably lower mass, on average, in the former case than the latter.
\end{itemize}

To explore the fate of the secondary accreted satellites, in Figs. \ref{fig:satTrackingA} and \ref{fig:satTrackingB}, we track satellites brought in by the two mergers A and B respectively for the \XXC, \XC{} and \XCXX{} simulations through their stellar particles (the two intermediate runs are omitted for brevity). Satellites are selected with the criteria described in Section \ref{sec:selection}, but now with respect to the merging system instead of the main progenitor. For clarity, we only track particles that are initially within $0.25\times \RGEN{halo}$ of the satellite as defined by the halo finder. Particles are coloured by the stellar mass rank (rather than stellar mass itself) of the satellite they originate from. In both figures, we show snapshots at the beginning and end of the respective merger (along with the beginning and end of merger B in Fig. \ref{fig:satTrackingA}) and 1 \& 2 Gyr after the end of merger B. The main host halo is shown by the black dashed circle in each panel, while the grey dashed circle indicates the relevant merging system.

Fig. \ref{fig:satTrackingA} shows that the majority of the satellites brought in by merger A have been disrupted by the beginning of merger B and very few if any merger A satellites survive during merger B. This accounts for the rapid drop in number of satellites seen towards the end of merger A in Fig. \ref{fig:nSatEvolutionWRTMerger}. The fact that mergers A and B occur in quick succession may also be a factor in the disruption of satellites from the former. The occurrence of multiple mergers occurring within a short time span is common at early epochs due to the nature of $\Lambda$CDM; in fact, the particular set of merger histories explored in our simulations were shown to be consistent with predicted merger rates in a $\Lambda$CDM universe \citep[see][]{Rey2023}. The haloes have relatively quiet merger histories after merger B by design. Hence, Fig. \ref{fig:satTrackingB} shows that the smaller number of satellites brought in by merger B have a higher probability of survival on 1-2~Gyr timescales. Exploring the impact of a concentrated merger history versus a more spread out one on our results is beyond the scope of these simulations, but is a topic that could be explored in the future through additional genetic modification of the merger timings.

The overall picture that emerges from examining these merger histories in detail is that the mass ratio of any merger cannot be fully separated from the overall environment that the system is embedded in. Note that the genetic modification algorithm constructs the closest possible sets of $\Lambda$CDM ICs subject to the desired change, and so give us a lower bound on how environmental factors become intertwined with merger constraints; the interrelationships will become even harder to separate in volume simulations or observations. Furthermore, while there are noticeable impacts of increasing/decreasing the mass ratio of the target merger as described here, the trends are not always monotonic w.r.t. to the GMs applied. This is evident in the initial increase of satellite numbers during merger A in Figs. \ref{fig:centralEvolMhaloMstarNsat}(b) and \ref{fig:nSatEvolutionWRTMerger}, where the \XCX{} simulation has more satellites than the \XC{} one, or in the satellite mass functions in Fig. \ref{fig:satMFAfterMerger}(c-e), where the \XXCV{} (\XCX{}) scenario may have more (fewer) satellites than the \XXC{} (\XCXX{}) one. 

\subsection{Summary of observational effects}

Our results show that the impact of varying the mass ratios of an early merger (i.e. several Gyrs ago) at a fixed $z=0$ halo mass is unlikely to be detectable on the satellite MFs at $z=0$. In future work, we will consider whether the satellite population retains a stronger memory of its merger history through additional factors such as the star-formation histories, quenched fractions and metallicities of the satellites. However, the merger histories do affect the satellite demographics in the system over timescales of up to $\sim 5$ Gyr, which implies that the \emph{recent} merger histories of host systems may be an important source of scatter in the total number of satellites and their mass functions. In our simulations, the timescale over which this effect is important is $\sim 5$ Gyr. 

The results of Fig. \ref{fig:nSatEvolutionMStar} also indicate that the differences in numbers of satellites are not simply the result of different central masses at a given cosmic time or time interval after a significant merger. Considering the reverse proposition, the recent merger histories may also be a significant source of uncertainty when using the (dynamical) properties of the satellites in reconstructing the properties of the central galaxy \citep[e.g. see][for techniques to constrain galaxy cluster masses]{Gifford2013,Armitage2019,Li2019}. As previously mentioned, direct comparisons with observational quantities requires careful consideration of selection effects and measurement biases, so we leave an investigation into the impact of our results on such techniques to a future paper. 

\section{Conclusions}   \label{sec:conclusions}
We have explored the impact of a galaxy's merger history on its population of dwarf satellites with the \textsc{vintergatan-gm} suite of simulations. \textsc{vintergatan-gm} is a set of five zoom-in simulations of a MW-mass system consisting of a fiducial simulation exhibiting a significant merger at $z \approx 2$, and four variations in which the ICs have been modified with the \textsc{GenetIC} algorithm to vary the mass ratio of this $z \approx 2$ merger, while attaining the same halo mass at $z=0$. This targeted modification allows us to isolate, to the maximum extent possible, the impact of varying the merger history of the host system on its satellite abundance and mass function. We summarize our main conclusions here.

\begin{itemize}
    \item The number and mass functions of satellites around the central galaxy \emph{at early times} are significantly impacted by the hosts' merger history; the smaller (larger) merger scenarios result in more (fewer) satellites being present after the end of the target merger. However, these differences are then compensated for at later times.
    \item The mass functions of the satellites are noticeably impacted by the target merger for $\sim 2.25-4.25$ Gyr after it ends (which corresponds to $\sim 2-4$ dynamical times for the system at the end of the targeted merger), with the smaller merger scenarios resulting in more low-mass ($\MSTAR\sim10^{6-7.5}\MSUN$) satellites compared to the larger merger scenarios.
    \item Modifying the \emph{early} merger history of the central galaxy has little impact on the total number of satellites surrounding it and their mass functions at $z=0$. This indicates that the satellite MFs may not retain the memory of mergers occurring at early times to present day, even though the mergers can significantly alter the properties of the central galaxy itself. Additional observables such as metallicities and quenched fractions will be examined in future work.
\end{itemize}

Our results indicate that the merger history of a galaxy can have noticeable impacts on its satellite population, but that any such impact on the stellar mass function is retained only for $\sim 2.25-4.25$ Gyr after a significant merger when considered at fixed eventual halo mass. Nonetheless, this is a cosmologically significant time window and therefore it may indeed be possible to link the recent merger activity of a galaxy to its satellite stellar mass function observationally. As previously emphasised by \cite{Rey2023}, the GM technique highlights that linking any particular observations directly to the consequences of a purported merger is risky in $\Lambda$CDM, due to its highly correlated structure. Our results further underscore this point, showing that effects on the satellite population arise for a number of reasons both directly and indirectly linked to a merger. A combination of GM simulations and large volume simulations is probably required to understand how to disentangle these effects for future observations. Although our results are produced for MW-mass systems, logically such trends are likely to be seen at all mass regimes. Finally, as mentioned earlier, while these results consider the impact of the mergers on the satellite mass functions, it is possible that the impact on other satellite properties such as quenched fractions and metallicity distributions may be more pronounced and long-lived. Our future work will focus on understanding the response to these properties along with incorporating observational selection effects in order to make robust comparisons between our simulations and observed MW dwarfs.

\section*{Author contributions}

GJ: Data curation, formal analysis, investigation, writing -- original draft. AP: Conceptualization, funding acquisition, methodology, resources, writing -- review \& editing. OA: Conceptualization, resources, writing -- review \& editing MR: Conceptualization, data curation, writing -- review \& editing. JR: Writing -- review \& editing FR: Writing -- review \& editing.

\section*{Acknowledgements}
We thank the anonymous referee for their insightful comments which were useful in improving the initial manuscript. This project has received funding from the European Union’s Horizon 2020 research and innovation programme under grant agreement No. 818085 GMGalaxies. This study used computing equipment funded by the Research Capital Investment Fund (RCIF) provided by UKRI, and partially funded by the UCL Cosmoparticle Initiative. OA and FR acknowledge support from the Knut and Alice Wallenberg Foundation, the Swedish Research Council (grant 2019–04659) and the Royal Physiographic Society in Lund. We acknowledge PRACE for awarding us access to Joliot-Curie at GENCI/CEA, France to perform the simulations presented in this work. Parts of the computations and data storage were enabled by resources (allocations SNIC 2022/5-136 and SNIC 2022/6-75) provided by the Swedish National Infrastructure for Computing (SNIC) at National Supercomputer Centre at Link\"oping University partially funded by the Swedish Research Council through grant agreement no. 2018-05973. MR is supported by the Beecroft Fellowship funded by Adrian Beecroft. FR acknowledges support provided by the University of Strasbourg Institute for Advanced Study (USIAS), within the French national programme Investment for the Future (Excellence Initiative) IdEx-Unistra. This work also made extensive use of \textsc{numpy} \citep{Harris2020Numpy} and \textsc{matplotlib} \citep{Hunter2007Matplotlib} packages.

\section*{Data Availability}

The data underlying this article will be shared upon reasonable request to the corresponding author.



\bibliographystyle{mnras}
\bibliography{PRACE_dwarf_MW_satellites} 







\bsp	
\label{lastpage}
\end{document}